\documentclass{aastex}
\usepackage{emulateapj5}
\usepackage{apjfonts}
\usepackage{epsf}
\slugcomment{UTAP-446}
\shorttitle{A NEW METHOD FOR CALCULATING UHECR ARRIVAL DISTRIBUTION}
\shortauthors{YOSHIGUCHI ET AL.}
\begin{document}
%%%%%%%%%%%%%%%%%%%%%%%%%%%%%%%%%%%%%%%%%%%%%%%%%%%%%%%%%%%%%%%%%%%%%%
%%%%%%%%%%%%%%%%%%%%%%%%%%%%%%%%%%%%%%%%%%%%%%%%%%%%%%%%%%%%%%%%%%%%%%
\title{A New Method for Calculating Arrival Distribution of Ultra-High
Energy Cosmic Rays above $10^{19}$ eV with Modifications by the
Galactic Magnetic Field}
%%%%%%%%%%%%%%%%%%%%%%%%%%%%%%%%%%%%%%%%%%%%%%%%%%%%%%%%%%%%%%%%%%%%%%
%%%%%%%%%%%%%%%%%%%%%%%%%%%%%%%%%%%%%%%%%%%%%%%%%%%%%%%%%%%%%%%%%%%%%%
%
%%%%%%%%%%%%%%%%%%%%%%%%%%%%%%%%%%%%%%%%%%%%%%%%%%%%%%%%%%%%%%%%%%%%%%
\author{Hiroyuki Yoshiguchi\altaffilmark{1}, Shigehiro
Nagataki\altaffilmark{1,2}, and Katsuhiko Sato\altaffilmark{1,2}}

\altaffiltext{1}{Department of Physics, School of Science, the University
of Tokyo, 7-3-1 Hongo, Bunkyoku, Tokyo 113-0033, Japan}
\altaffiltext{2}{Research Center for the Early Universe, School of
Science, the University of Tokyo,
7-3-1 Hongo, Bunkyoku, Tokyo 113-0033, Japan}

\email{hiroyuki@utap.phys.s.u-tokyo.ac.jp}
%%%%%%%%%%%%%%%%%%%%%%%%%%%%%%%%%%%%%%%%%%%%%%%%%%%%%%%%%%%%%%%%%%%%%%
%
\received{}
\accepted{}
\begin{abstract}
We present a new method for calculating arrival distribution
of Ultra-High Energy Cosmic Rays (UHECRs) including modifications by
the galactic magnetic field.
We perform numerical simulations of UHE anti-protons, which are
injected isotropically at the earth, in the Galaxy and record the
directions of velocities at the earth and outside the Galaxy for all
of the trajectories.
We then select some of them so that the resultant mapping of
the velocity directions outside the Galaxy of the selected
trajectories corresponds to a given source location scenario, applying
Liouville's theorem.
We also consider energy loss processes of UHE protons in the
intergalactic space.
Applying this method to our source location scenario which is adopted
in our recent study and can explain the AGASA observation above $4
\times 10^{19}$ eV, we calculate the arrival distribution of UHECRs
including lower energy ($E>10^{19} {\rm eV}$) ones.
We find that our source model can reproduce the large-scale isotropy
and the small-scale anisotropy on UHECR arrival distribution above
$10^{19}$ eV observed by the AGASA.
We also demonstrate the UHECR arrival distribution above $10^{19}$ eV
with the event number expected by future experiments in the next few
years.
The interesting feature of the resultant arrival distribution is the
arrangement of the clustered events in the order of their energies,
reflecting the directions of the galactic magnetic field.
This is also pointed out by \cite{stanev02}.
This feature will allow us to obtain some kind of information about
the composition of UHECRs and the magnetic field with increasing
amount of data.
\end{abstract} 
%
%%%%%%%%%%%%%%%%%%%%%%%%%%%%%%%%%%%%%%%%%%%%%%%%%%%%%%%%%%%
%%%%%%%%%%%%%%%%%%%%%%%%%%%%%%%%%%%%%%%%%%%%%%%%%%%%%%%%%%%
\section{INTRODUCTION} \label{intro}
%%%%%%%%%%%%%%%%%%%%%%%%%%%%%%%%%%%%%%%%%%%%%%%%%%%%%%%%%%%
%%%%%%%%%%%%%%%%%%%%%%%%%%%%%%%%%%%%%%%%%%%%%%%%%%%%%%%%%%%

There is no statistically significant large scale anisotropy
in the observed arrival distribution of ultra-high energy cosmic rays
(UHECRs) above $10^{19}$ eV \citep{takeda99}.
This may imply an extragalactic origin of
cosmic rays above $10^{19}$ eV, combined with the change of spectral
slope of the observed energy spectrum at $\sim 10^{19}$ eV
\citep{bird94,yoshida95,takeda98}.
Another important feature of the UHECR arrival distribution is the
small scale clusterings of the arrival directions
\citep{takeda99,takeda01}.
The current AGASA data set of 57 events above 4 $\times 10^{19}$ eV
contains four doublets and one triplet within
a separation angle of 2.5$^\circ$.
Chance probability to observe such clusters under an isotropic
distribution is only about 1 $\%$ \citep*{hayashida00,takeda01}.

On the other hand, the cosmic-ray energy spectrum does not show the
GZK cutoff \citep*{greisen66,zatsepin66} because of
photopion production with the photons of the cosmic microwave
background (CMB) and extends above $10^{20}$ eV
\citep*{takeda98}.
The discrepancy between the AGASA and the High Resolution Fly's Eye
\citep*[HiRes; ][]{wilkinson99}, which reports the cosmic ray flux
with the GZK cut-off around $10^{20}$ eV \citep*{abu02}, remains to be
one of the major open question in astroparticle physics.
This issue is left for future investigation by new large-aperture
detectors under development, such as South and North Auger project
\citep*{capelle98}, the EUSO \citep*{euso92}, and the OWL
\citep*{owl00} experiments.

In our recent work \citep*[][hereafter Paper I]{yoshiguchi03a}, we
perform numerical simulations for propagation of UHE protons in
intergalactic space, and examine whether the present AGASA observation
above $4 \times 10^{19}$ eV can be explained by a bottom-up scenario
in which the source distribution of UHECRs is proportional to that of
galaxies.
We use the Optical Redshift Survey \citep*[ORS; ][]{santiago95} to
construct realistic source models of UHECRs.

In Paper I, we calculate both the energy spectrum and arrival
directions of UHE protons, and compare the results with the AGASA
observation above $4 \times 10^{19}$ eV.
We find that the large-scale isotropy and the small-scale anisotropy
of the UHECR arrival distribution observed by the AGASA can be
reproduced when $\sim 1/50$ of the ORS sample more luminous than
$-20.5$ mag are selected as UHECR sources, in the case of weak
extragalactic magnetic field (EGMF $B \le 1$ nG).
In terms of the source number density, this constraint
corresponds to $\sim 10^{-6}$ Mpc$^{-3}$.

%This is because luminous galaxies in the Local Super Cluster (LSC)
%are distributed more widely than faint galaxies, contrary to general
%clusters of galaxies \citep*{yoshiguchi03b}.

The small scale anisotropy can not be well reproduced in the case of
strong EGMF ($B \ge 10$ nG), because the correlation at small scale
between events which originate from a single source is eliminated, or
the correlation continues to larger angle scale, due to large
deflection when UHECRs propagate in the EGMF from sources to the
earth.
Although \cite{isola02} and \cite{sigl03} conclude that the expected
small-scale anisotropy and large-scale isotropy for local enhancement
of UHECR sources in the LSC in the presence of the strong EGMF ($\sim
1 \mu$ G) are in marginal agreement with the AGASA, the consistency is
somewhat worse than that predicted by our scenario for $B=1$ nG.
Of course, we can not draw any firm conclusion about the strength of
the EGMF, considering the current limited amount of data.
However, we assume extremely weak EGMF throughout the paper.

If local enhancement of UHECR sources in the LSC
\citep*{isola02,sigl03} is disfavored from the observations, there is
no way that explains the observed extension
of the cosmic-ray spectrum beyond the GZK cutoff.
Our conclusion in Paper I is that a large fraction of cosmic rays above
$10^{20}$ eV observed by the AGASA experiment might originate in the
top-down scenarios, or that the energy spectrum measured by the Hires
experiment might be better.

As mentioned above, we obtain the constraint on the source number
density as $\sim 10^{-6}$ Mpc$^{-3}$ by comparing our model prediction
with the AGASA data only above $4 \times 10^{19}$ eV.
It is very important to examine whether our source model can explain
the AGASA data including lower energy $(\sim 10^{19} {\rm eV})$ one.
On the other hand, the arrival directions of UHECRs above $10^{19}$ eV
are modified by the galactic magnetic field (GMF) by a few $- \sim 10$
degrees.
In order to accurately calculate the expected UHECR arrival
distribution and compare with the observations, the effect of the GMF
should be taken into account.

The first step of the studies of UHECR propagation in the GMF is found
in \cite{stanev02}.
\cite{stanev02} calculate the expected arrival distribution of
UHECRs above $10^{19.4}$ eV for several source location scenarios.
They perform numerical simulations of UHECR propagation in the Galaxy
injected from sources toward the earth.
The radius of the earth (detector) must be so small that the
unavoidable smearing in arrival angle is kept smaller than the
accuracy of arrival direction determination $\sim$ a few degree
\citep{takeda01}.
In this case, the number fraction of injected UHECRs arriving at the
earth is very small.
This requires a large number of particles to be propagated, which
takes enormous CPU time.

In this paper, we present a new method for calculating UHECR arrival
distribution which can be applied to several source location scenarios
including modifications by the GMF.
We numerically calculate the propagation of anti-protons from the
earth toward the outside of the Galaxy (in this study, we set a
sphere centered around the Galactic center with radius $r_{\rm src}=$40
kpc as the boundary condition), including the effects of Lorentz force
due to the GMF.
The anti-protons are ejected isotropically from the earth.
By this calculation, we can obtain the trajectories and the sky map of
position of anti-protons that have reached to the boundary at radius
$r_{\rm src}=$ 40 kpc.

Next, we regard the obtained trajectories as the ones of PROTONs from
the outside of the galaxy toward the earth.
Also, we regard the obtained sky map of position of anti-proton at the
boundary as relative probability distribution (per steradian) for
PROTONs to be able to reach to the earth for the case in which the
flux of the UHE protons from the extra-galactic region is isotropic
(in this study, this flux corresponds to the one at the boundary
$r_{\rm src}=40$ kpc).
This treatment is supported by the Liouville's theorem.
When the flux of the UHE protons at the boundary is anisotropic (e.g.,
the source distribution is not isotropic), this effect can be included
by multiplying this effect (that is, by multiplying the probability
density of arrival direction of UHE protons from the extra-galactic
region at the boundary) to the obtained relative probability
density distribution mentioned above.

By adopting this new method, we can consider only the trajectories of
protons which arrive to the earth, which, of course, helps us to save
the CPU time efficiently and makes calculation of propagation of CRs
even with low energies ($\sim 10^{19}$ eV) possible within a
reasonable time.
We also consider the energy loss processes of UHE protons in the
intergalactic space, which is not taken into account by \cite{stanev02}.

With this method, we calculate the UHECR arrival distribution above
$10^{19}$ eV for our source scenario which can explain the current
AGASA observation above $10^{19.6}$ eV.
Using the harmonic amplitude and the two point correlation function as
statistical quantities, we compare our model prediction with the AGASA
observation.
We also demonstrate the arrival distribution of UHECRs with the event
number expected by future experiments such as South and North Auger
project \citep*{capelle98}, the EUSO \citep*{euso92}, and the OWL
\citep*{owl00} experiments.

In section~\ref{gmf}, we introduce the GMF model.
We explain the method for calculating UHECR arrival distribution in
section~\ref{method}.
Results are shown in section~\ref{result}.
In section~\ref{summary}, we summarize the main results.

%%%%%%%%%%%%%%%%%%%%%%%%%%%%%%%%%%%%%%%%%%%%%
\section{GALACTIC MAGNETIC FIELD} \label{gmf}
%%%%%%%%%%%%%%%%%%%%%%%%%%%%%%%%%%%%%%%%%%%%%
In this study, we adopt the GMF model used by \cite{stanev02},
which is composed of the spiral and the dipole field.
In the following, we briefly introduce this GMF model.

Faraday rotation measurements indicate that the GMF in the
disk of the Galaxy has a spiral structure with field reversals
at the optical Galactic arms \citep{beck01}.
We use a bisymmetric spiral field (BSS) model, which is favored
from recent work \citep{han99,han01}.
The Solar System is located at a distance
$r_{\vert\vert}=R_\oplus=8.5$ kpc from the center of the Galaxy in the
Galactic plane.
The local regular magnetic field in the vicinity of the Solar System
is assumed to be 
$B_{\rm Solar} \sim 1.5~\mu{\rm G}$ in the direction $l=90^{\rm o}+p$
where the pitch angle is $p=-10^{\rm o}$ \citep{han94}.

In the polar coordinates $(r_{\vert\vert},\phi)$,
the strength of the spiral field in the Galactic plane is given by
\begin{equation}
B(r_{\vert\vert},\phi)=
B_0~\left({R_\oplus \over r_{\vert\vert}}\right)~
\cos\left(\phi - \beta \ln {r_{\vert\vert}\over r_0} \right)
\end{equation}  
where $B_0=4.4~\mu$G, $r_0= 10.55$ kpc and $\beta=1/\tan p=-5.67$.
The field decreases with Galactocentric distance as $1/r_{\vert\vert}$ 
and it is zero for $r_{\vert\vert}>20$ kpc. 
In the region around the Galactic center ($r_{\vert\vert} < 4$ kpc) 
the field is highly uncertain, and thus assumed to be constant and
equal to its value at $r_{\vert\vert}=4$ kpc.

The spiral field strengths above and below the Galactic plane are taken to
decrease exponentially with two scale heights \citep{stanev96}
\begin{equation}
\vert B(r_{\vert\vert},\phi,z)\vert = 
\vert B(r_{\vert\vert},\phi)\vert 
\left\{
\begin{array}{lcl}
\exp(-z) :  & \vert z \vert \leq 0.5~ {\rm kpc} \\ 
\exp( \frac{-3}{8})~\exp(\frac{-z}{4}) :  & \vert z \vert > 0.5~ {\rm kpc} 
\end{array}
\right.
\label{eq:b-height}
\end{equation} 
where the factor $\exp(-3/8)$ makes the field continuous in $z$.
The BSS spiral field we use is of even parity, that is,
the field direction is preserved at disk crossing.

Observations show that 
the field in the Galactic halo is much weaker than that in the disk.
In this work we assume that the regular field corresponds to a 
A0 dipole field as suggested in \citep{han02}.
In spherical coordinates $(r,\theta,\varphi)$ 
the $(x,y,z)$ components of the halo field are given by:
\begin{eqnarray}
B_x=-3~\mu_{\rm G}~{\sin\theta \cos\theta \cos\varphi}/
r^3 \nonumber \\ 
B_y=-3~\mu_{\rm G}~{\sin\theta \cos\theta \sin\varphi}/
r^3 \\
B_z=\mu_{\rm G}~{(1-3\cos^2\theta)}/r^3 ~~~~~~~ \nonumber
\label{eq:bdipole}
\end{eqnarray}
where $\mu_{\rm G}\sim 184.2~{\rm \mu G~kpc^3}$ is the magnetic moment
of the Galactic dipole.
The dipole field is very strong in the central region 
of the Galaxy, but is only 0.3 $\mu$G in the vicinity
of the Solar system, directed toward the North Galactic Pole.

There is a significant turbulent component, $B_{\rm ran}$,
of the Galactic magnetic field.
Its field strength is difficult to measure and results found in
literature are in the range of $B_{\rm ran} = 0.5 \dots 2 B_{\rm reg}$
\citep{beck01}.
However, we neglect the random field throughout the paper, in order to
make easy to see the effects of the regular field, such as the
arrangement of the clustered event in the order of their energies
(section~\ref{arrival_future}).
Possible dependence of the results on the random field is discussed in
the section~\ref{statistics_arrival}.

%%%%%%%%%%%%%%%%%%%%%%%%%%%%%%%%%%%%%%%%%%%%%
\section{NUMERICAL METHOD} \label{method}
%%%%%%%%%%%%%%%%%%%%%%%%%%%%%%%%%%%%%%%%%%%%%
%-----------------------------------------------------------------
\subsection{Propagation of UHECRs in the Intergalactic Space}
\label{sim}
%-----------------------------------------------------------------

The energy spectrum of UHECRs injected at extragalactic sources is
modified by the energy loss processes when they propagate in the
intergalactic space. 
This subsection provides the method of Monte Carlo simulations
for propagation of UHE protons in intergalactic space.

We assume that UHECRs are protons injected with a power law
spectrum within the range of ($10^{19}$ - $10^{22}$)eV.
10000 protons are injected in each of 31 energy bins, that is,
10 bins per decade of energy.
Then, UHE protons are propagated including the energy loss processes
(explained below) over $3$ Gpc for $15$ Gyr.
We take a power law index as 2.6 in order to fit the calculated energy
spectrum to the one observed by the AGASA \citep{marco03}.

UHE protons below $\sim 8 \times 10^{19}$ eV lose their energies
mainly by pair creations and adiabatic energy losses, and above it by
photopion production \citep*{berezinsky88,yoshida93} in collisions
with photons of the CMB.
We treat the adiabatic loss as a continuous loss process.
We calculate the redshift $z$ of source at a given distance using the
cosmological parameters $H_0=71$ km s$^{-1}$ Mpc$^{-1}$,
$\Omega_m=0.27$, and $\Omega_{\Lambda}=0.73$.
Similarly, the pair production can be treated as a continuous loss process
considering its small inelasticity ($\sim 10^{-3}$).
We adopt the analytical fit functions given by \cite{chodorowski92}
to calculate the energy loss rate for the pair production
on isotropic photons.
The same approach has been adopted in our previous studies
\citep*[Paper I, ][]{ide01,yoshiguchi03c}.

On the other hand, protons lose a large fraction of their energy in
the photopion production.
For this reason, its treatment is very important.
We use the interaction length and the energy distribution of
final protons as a function of initial proton energy
which is calculated by simulating the photopion
production with the event generator SOPHIA \citep*{sophia00}.

In this study, we neglect the effect of the EGMF because of the
following two reasons.
First, numerical simulations of UHECR propagation in the EGMF
including lower energy ($\sim 10^{19}$ eV) ones take a long CPU time.
Second, we show in our previous study that small scale clustering can
be well reproduced in the case of weak EGMF $(B < 1{\rm nG})$
(Paper I).
\cite{isola02} and \cite{sigl03} show that the expected small scale
anisotropy for local enhancement scenario of UHECR sources in the
presence of strong EGMF ($\sim 1 \mu$ G) in the Local Super Cluster is
marginally consistent with the AGASA observation.
However, the consistency of small-scale anisotropy is somewhat worse
than that predicted by our scenario in the case of weak EGMF
(Paper I).
Although we can not draw any firm conclusion because of the limited
amount of data, we assume extremely weak EGMF throughout the paper.

%-----------------------------------------------------------------
\subsection{Source Distribution}
\label{method_source}
%-----------------------------------------------------------------

In this study, we apply the method for calculating the UHECR arrival
distribution with modifications by the GMF
(section~\ref{calc_arrival}) to our source location scenario, which is
constructed by using the ORS \citep*{santiago95} galaxy catalog.
As mentioned in section~\ref{intro}, we show in Paper I that the
arrival distribution of UHECRs observed by the AGASA can be reproduced
when $\sim 1/50$ of the ORS galaxies more luminous than $M_{\rm
lim}=-20.5$ is selected as UHECR sources.
We consider only this source model throughout the paper.
It is unknown how much an ultimate UHECR source
contribute to the observed cosmic ray flux.
In paper I, we thus consider the two cases in which (1) all galaxies
inject the same amount of cosmic rays, or (2) they inject cosmic rays
proportionally to their absolute luminosity.
However, we find that the results in the two cases do not differ from
each other, as far as we focus on the luminous galaxies as UHECR
sources.
Accordingly, we restrict ourselves to the case that all galaxies
inject the same amount of cosmic rays.

In order to calculate the energy spectrum and the distribution of arrival
directions of UHECRs realistically, there are two key elements
of the galaxy sample to be corrected.
First, galaxies in a given magnitude-limited sample are biased tracers
of matter distribution because of the flux limit \citep*{yoshiguchi03b}.
Although the sample of galaxies more luminous than $-20.5$ mag is
complete within 80 $h^{-1}$ Mpc (where $h$ is the Hubble constant
divided by 100 km s$^{-1}$ and we use $h=0.71$), it does not contain
galaxies outside it for the reason of the selection effect.
We distribute sources of UHECRs outside 80 $h^{-1}$ Mpc homogeneously.
Their number density is set to be equal to that inside 80 $h^{-1}$
Mpc.
We do not take into account luminosity evolution for simplicity.

Second, our ORS sample does not include galaxies
in the zone of avoidance ($|b|<20^{\circ}$).
In the same way, we distribute UHECR sources in this region homogeneously,
and calculate its number density from the number of galaxies in the
observed region.

%-----------------------------------------------------------------
\subsection{Calculation of the UHECR Arrival Distribution with
modifications by the GMF}
\label{calc_arrival}
%-----------------------------------------------------------------

In this subsection, we present the method of calculation of UHECR
arrival distribution with modifications by the GMF.
We start by injecting anti-protons from the earth isotropically, and
follow each trajectory until

1. anti-proton reaches a sphere of radius $40$ kpc centered at the galactic
   center, or

2. the total path length traveled by anti-proton is larger than $200$
   kpc,

by integrating the equations of motion in the magnetic field.
It is noted that we regard these anti-protons as PROTONs injected from
the outside of the Galaxy toward the earth.
The number of propagated anti-proton is 2,000,000.
We have checked that the number of trajectories which are stopped by
the limit (2) is smaller than $0.1$\% of the total number.
The energy loss of protons can be neglected for these distances.
Accordingly, we inject anti-protons with injection spectrum similar to
the observed one $\sim E^{-2.7}$.
(Note that this is not the energy spectrum injected at extragalactic
sources.)

\begin{figure*}
\begin{center}
%\leavevmode
\epsscale{2.0}
\plotone{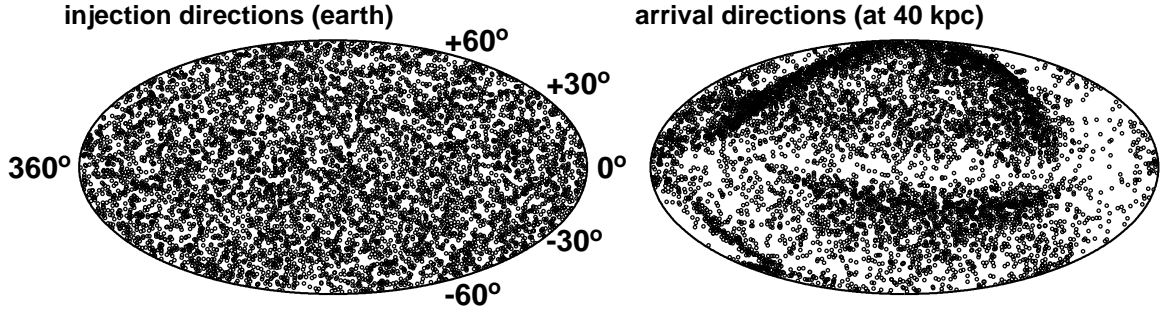} 
\caption{
Arrival directions of anti-protons with $E>10^{19.0}$
eV at the sphere of Galactocentric radius of $40$ kpc (right panel)
in the galactic coordinate.
The anti-protons are injected at the earth isotropically (as shown in
the left panel) with an injection spectrum $E^{-2.7}$.
\label{event}}
\end{center}
\end{figure*}

The result of the velocity directions of anti-protons at the sphere of
radius $40$ kpc is shown in the right panel of figure~\ref{event} in
the galactic coordinate.
From Liouville's theorem, if the cosmic-ray flux outside the Galaxy is
isotropic, one expects an isotropic flux at the earth even in the
presence of the GMF.
This theorem is confirmed by numerical calculations shown in figure 6
of \cite{stanev02}, which is the same figure as our figure~\ref{event}
except for threshold energy.
Thus, the mapping of the velocity directions in the right panel of
figure~\ref{event} corresponds to the sources which actually give rise
to the flux at the earth in the case that the sources (including ones
which do not actually give rise to the flux at the earth) are
distributed uniformly.

We calculate the UHECR arrival distribution for our source scenario
using the numerical data of the propagation of UHE anti-protons in the
Galaxy.
Detailed method is as follows.
At first, we divide the sky into a number of bins with the same solid
angle.
The number of bins is taken to be $360 (l) \times 200 (b)$.
We then distribute all the trajectories into each bin according to
their directions of velocities (source directions) at the sphere of
radius $40$ kpc.
Finally, we randomly select trajectories from each bin with
probability $P_{\rm selec} (j,k,E)$ defined as
\begin{equation}
P_{\rm selec} (j,k,E) \propto \sum _{i} \frac{1}{d_i^2} \, 
\frac{dN/dE(d_i,E)}{E^{-2.7}}.
\label{def_pselec} 
\end{equation}
Here subscripts j and k distinguish each cell of the sky, $d_i$ is
distance of each galaxy within the cell of (j,k), and the summation
runs over all of the galaxies within it.
$E$ is the energy of proton, and $dN/dE(d_i,E)$ is the energy
spectrum of protons at our galaxy injected at a source of distance
$d_i$.

The normalization of $P_{\rm selec} (j,k,E)$ is determined so as to
set the total number of events equal to a given number, for example,
the event number of the current AGASA data.
When $P_{\rm selec}>1$, we newly generate events with number of
$(P_{\rm selec}-1) \times N(j,k)$, where $N(j,k)$ is the number of
trajectories within the sky cell of $(j,k)$.
The arrival angle of newly generated proton (equivalently, injection
angle of anti-proton) at the earth is calculated by adding a normally
distributed deviate with zero mean and variance equal to the
experimental resolution $2.8^{\circ}$ $(1.8^{\circ})$ for $E>10^{19}$
eV $(4 \times 10^{19} {\rm eV})$ to the original arrival angle.
We perform this event generation 20 times in order to calculate the
averages and variances, due to the finite number of the simulated
events, of the statistical quantities (section~\ref{statistics}).

%-------------------------------------------------------------------------
\subsection{Statistical Methods}\label{statistics}
%-------------------------------------------------------------------------

In this subsection, we explain the two statistical quantities,
the harmonics analysis for large scale anisotropy \citep*{hayashida99},
the two point correlation function for small scale anisotropy.

The harmonic analysis to the right ascension distribution of events
is the conventional method to search for global anisotropy
of cosmic ray arrival distribution.
For a ground-based detector like the AGASA, the almost
uniform observation in right ascension is expected.
The $m$-th harmonic amplitude $r$ is determined by fitting the distribution
to a sine wave with period $2 \pi /m$.
For a sample of $n$ measurements of phase,
$\phi_1$, $\phi_2$, $\cdot \cdot \cdot$, $\phi_n$
(0 $\le \phi_i \le 2 \pi$), it is expressed as
\begin{equation}
r = (a^2 + b^2)^{1/2}
\label{eqn121}
\end{equation}
where, $a = \frac{2}{n} \Sigma_{i = 1}^{n} \cos m \phi_i  $,
$b = \frac{2}{n} \Sigma_{i = 1}^{n} \sin m \phi_i  $.
We calculate the harmonic amplitude for $m=1,2$ from a set of events
generated by the method explained in the section~\ref{calc_arrival}.

If events with
total number $n$ are uniformly distributed in right ascension, the chance
probability of observing the amplitude $\ge r$ is given by,
\begin{equation}
P = \exp (-k),
\label{eqn13}
\end{equation}
where
\begin{equation}
k = n r^2/4.
\label{eqn14}
\end{equation}
The current AGASA 775 events above $10^{19}$ eV is consistent with
isotropic source distribution within 90 $\%$ confidence level
\citep*{takeda01}.
We therefore compare the harmonic amplitude for $P = 0.1$ with the
model prediction.

The two point correlation function $N(\theta)$ contains
information on the small scale anisotropy.
We start from a set of generated events.
For each event, we divide the sphere into concentric bins of
angular size $\Delta \theta$, and count the number of events falling
into each bin.
We then divide it by the solid angle of the corresponding bin,
that is,
\begin{eqnarray}
N ( \theta ) = \frac{1}{2 \pi | \cos \theta  - \cos (\theta + \Delta \theta)
|} \sum_{ \theta
\le  \phi \le \theta + \Delta \theta }  1 \;\;\; [ \rm  sr ^{-1} ],
\label{eqn100}
\end{eqnarray}
where $\phi$ denotes the separation angle of the two events.
$\Delta \theta$ is taken to be $1^{\circ}$ in this analysis.
The AGASA data shows correlation at small angle $(\sim 3^{\circ})$
with 2.3 (4.6) $\sigma$ significance of deviation from an isotropic
distribution for $E>10^{19}$ eV $(E>4 \times 10^{19} {\rm eV})$
\citep*{takeda01}.

%%%%%%%%%%%%%%%%%%%%%%%%%%%%%%%%%%%%%%%%%
\section{RESULTS} \label{result}
%%%%%%%%%%%%%%%%%%%%%%%%%%%%%%%%%%%%%%%%%

%-------------------------------------------------------------------------
\subsection{Arrival Distribution of UHECRs above $10^{19}$ eV}
\label{arrival}
%-------------------------------------------------------------------------

\begin{figure*}
\begin{center}
%\leavevmode
\epsscale{1.3}
\plotone{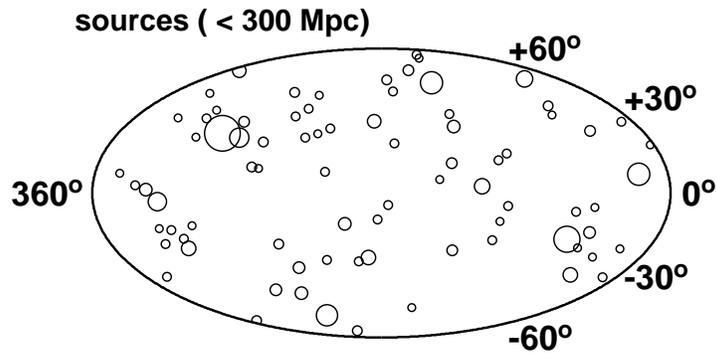} 
\caption{
Distribution of the sources in our model in the galactic coordinate.
We show only the sources within $300$ Mpc from us as circles of radius
inversely proportional to their distances.
It is noted that the sources outside 113$(=80 h^{-1})$ Mpc are
randomly distributed because the ORS sample does not contain any
galaxy outside it.
\label{source}}
\end{center}
\end{figure*}

In this subsection, we present the results of the arrival distribution
of UHECRs above $10^{19}$ eV, using the method explained in the
section~\ref{calc_arrival}.
At first, figure~\ref{source} shows the distribution of the sources
for a specific source selection when $\sim 1/50$ of the ORS galaxies
more luminous than $M_{\rm{lim}}=-20.5$ is randomly selected as UHECR
sources in the galactic coordinate.
We show only the sources within $300$ Mpc from us for clarity, as
circles of radius inversely proportional to their distances.
It is noted that the sources outside 113$(=80 h^{-1})$ Mpc are
randomly distributed because the ORS sample does not contain any
galaxy outside it.

%\vspace{0.5cm}
%%%%%%%%%%%%%%%%%%%%%%%%%%%%%%%%%%%%%%%%%%%%%%%%%%%%%%%%%%%%%%%
%\centerline{{\vbox{\epsfxsize=8.0cm\epsfbox{source.ps}}}}
%\figcaption{
%Distribution of the sources in our model in the galactic coordinate.
%We show only the sources within $300$ Mpc from us as circles of radius
%inversely proportional to their distances.
%It is noted that the sources outside 113$(=80 h^{-1})$ Mpc are
%randomly distributed because the ORS sample does not contain any
%galaxy outside it.
%\label{source}}
%\vspace{0.5cm}
%%%%%%%%%%%%%%%%%%%%%%%%%%%%%%%%%%%%%%%%%%%%%%%%%%%%%%%%%%%%%%%

\vspace{0.5cm}
%%%%%%%%%%%%%%%%%%%%%%%%%%%%%%%%%%%%%%%%%%%%%%%%%%%%%%%%%%%%%%%
\centerline{{\vbox{\epsfxsize=8.0cm\epsfbox{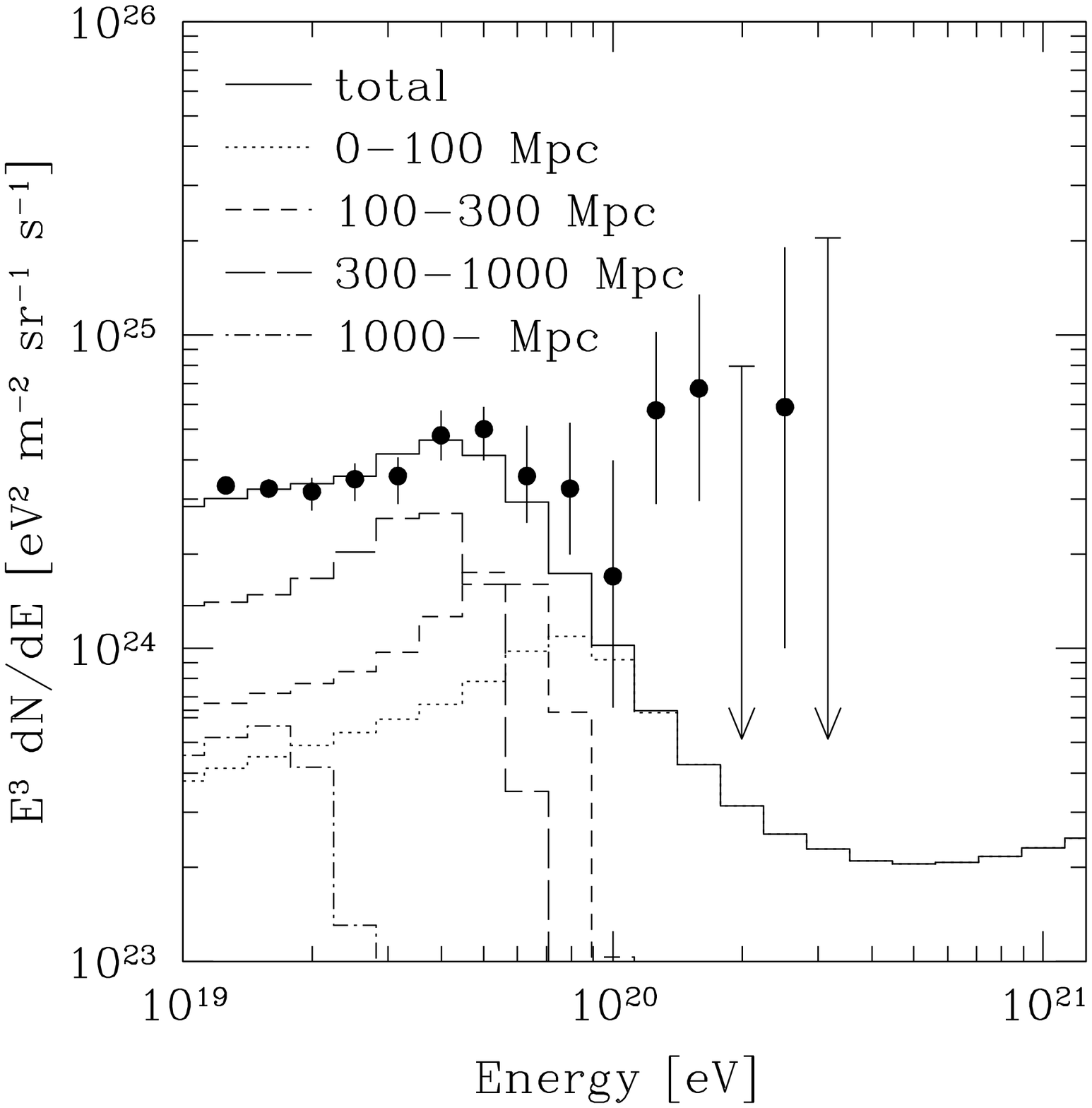}}}}
\figcaption{
Energy spectrum with injection spectrum $E^{-2.6}$,
predicted by the source model of figure~\ref{source}.
The contributions from sources at different distances are also shown.
We also show the observed cosmic-ray spectrum by the AGASA experiments
\citep*{hayashida00}.
\label{esp}}
\vspace{0.5cm}
%%%%%%%%%%%%%%%%%%%%%%%%%%%%%%%%%%%%%%%%%%%%%%%%%%%%%%%%%%%%%%%

We show in figure~\ref{esp} the expected energy spectrum for the
source model of figure~\ref{source}.
The injection spectrum is set to be $E^{-2.6}$.
The contributions from sources at different distances are also shown.
We also show the observed cosmic-ray spectrum by the AGASA experiments
\citep*{hayashida00}.
The resultant spectrum is in good agreement with the one observed by
the AGASA, except for $E>10^{20}$ eV.
As mentioned above, we conclude in Paper I that a large fraction of
cosmic rays above $10^{20}$ eV observed by the AGASA experiment might
originate in the top-down scenarios.
Accordingly, we consider only cosmic rays with $E<10^{20}$ eV
throughout the paper.

As mentioned above, \cite{stanev02} does not take the energy loss
processes in the intergalactic space into account.
Thus, they can not include the effects of difference between resultant
energy spectra injected at different distances into numerical
calculations.
In our calculations, however, sources at larger distance mainly
contribute to the cosmic ray flux at lower energies as is evident from
figure~\ref{esp}.
This enables us to calculate the arrival distribution of UHECRs under
more realistic situations.

Given the source distribution and the resultant energy spectrum as a
function of the source distance, we can calculate the right hand side
of Eq.~\ref{def_pselec}.
Then we perform the selection of trajectories according to the
probability $P_{\rm selec}$, as explained in the
section~\ref{calc_arrival}.

\begin{figure*}
\begin{center}
%\leavevmode
\epsscale{2.0}
\plotone{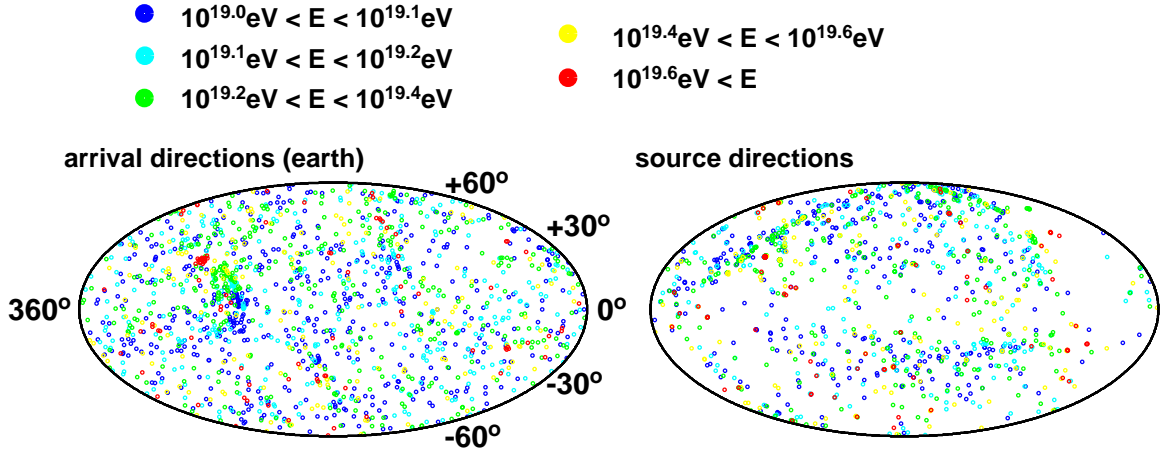} 
%\plotone{source.ps} 
\caption{
Arrival directions of protons with $E>10^{19.0}$ eV at the
earth (left panel) expected for the source model of
figure~\ref{source} in the galactic coordinate.
The events are shown by color according to their energies.
The event number within $-10^{\circ} \le \delta \le 80^{\circ}$ is set to be
equal to the one observed by the AGASA \citep*[775]{takeda01}
(The total number of events is $\sim 1500$).
The right panel is the mapping of the sources which actually give rise
to events shown in the left panel.
Note that this mapping differ from the distribution of the sources
shown in figure~\ref{source}.
\label{selected_event}}
\end{center}
\end{figure*}

One realization of the event generations is shown in
figure~\ref{selected_event}.
The events are shown by color according to their energies.
This figure corresponds to figure~\ref{event}.
That is, the injection directions of anti-protons at the earth
(figure~\ref{event}, left) corresponds to the arrival directions of
protons (figure~\ref{selected_event}, left).
Similarly, the arrival directions of anti-protons at the sphere of
Galactocentric radius of $40$ kpc (figure~\ref{event}, right) does to
the directions of the sources which actually give rise to the
cosmic-ray flux (figure~\ref{selected_event}, right).

For the source model of figure~\ref{source}, the nearest source is
located at $(b,l)=(31^{\circ},284^{\circ})$ and 64 Mpc from us.
A number of the simulated events are clustered at this direction as
seen in figure~\ref{selected_event}.
Furthermore, these events are aligned in the sky according to the
order of their energies, reflecting the direction of the GMF at this
direction.
As we will show in the section~\ref{arrival_future}, this interesting
feature of the UHECR arrival distribution becomes evident with
increasing amount of the event number.

%-------------------------------------------------------------------------
\subsection{Statistics on the UHECR Arrival Distribution}
\label{statistics_arrival}
%-------------------------------------------------------------------------

In this subsection, we show the results of the statistical quantities
on the UHECR arrival distribution above $10^{19}$ eV.
In the last section, we showed the results for a specific source
scenario when $\sim 1/50$ of the ORS galaxies more luminous than
$M_{\rm{lim}}=-20.5$ is randomly selected as UHECR sources.
However, the statistical quantities presented in this section are
calculated with not only the statistical error but also the variation
between different selections of source from our ORS sample.

\begin{figure*}
\begin{center}
%\leavevmode
\epsscale{1.2} 
\plotone{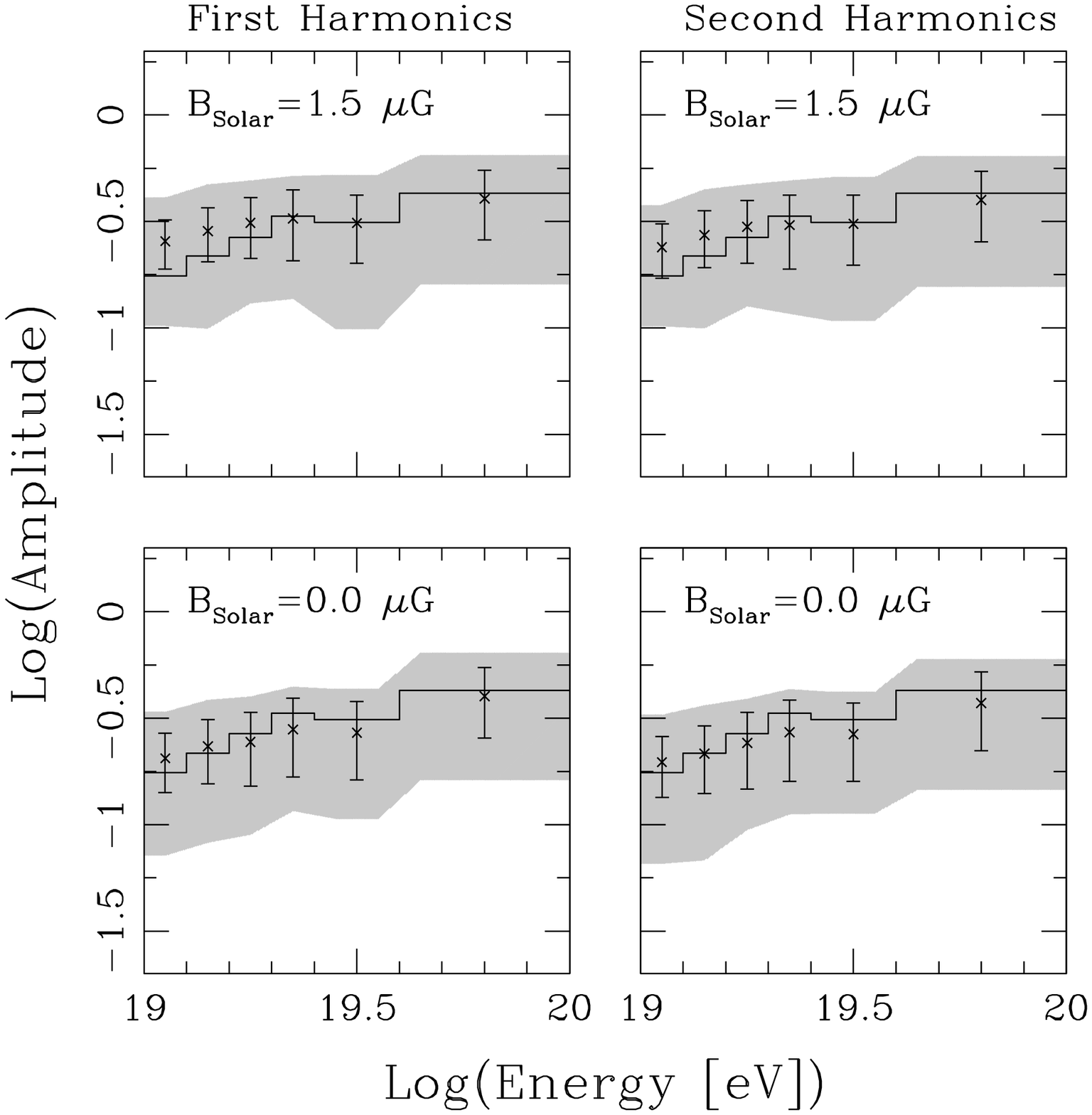} 
\caption{
Harmonic amplitude predicted by our source model as a function of the
cosmic-ray energies.
The errorbars represent the statistical fluctuations due to the finite
number of the simulated events, which is set to be equal to that
observed by the AGASA within $-10^{\circ} \le \delta \le 80^{\circ}$.
The shaded regions represent 1 $\sigma$ total error due to not only
the statistical error but also the source selections from our ORS
sample.
The region below the solid line is expected values due to the
statistical fluctuation of isotropic source distribution with the
chance probability larger than 10$\%$.
\label{amp}}
\end{center}
\end{figure*}

The upper panels of figure~\ref{amp} shows the first and the second
harmonics predicted by our source model as a function of the
cosmic-ray energies for $B_{\rm Solar}=1.5 \mu$ G, where $B_{\rm Solar}$ is
the strength of the GMF in the vicinity of the Solar system.
It is noted that we calculate the harmonic amplitudes for the
simulated events within only $-10^{\circ} \le \delta \le 80^{\circ}$
in order to compare with AGASA data.
The errorbars represent the statistical fluctuations due to the finite
number of the simulated events, which is set to be equal to that
observed by the AGASA (775 events for $E>10^{19}$ eV).
The event selections are performed $20$ times.
The shaded regions represent 1 $\sigma$ total error due to not only
the statistical error but also the source selections from our ORS
sample.
The random source selections are performed $100$ times.
The region below the histogram is expected values for the
statistical fluctuation of isotropic source distribution with the
chance probability larger than 10$\%$.

It is clear that our source model predicts the large-scale isotropy
fully consistent with that expected by uniform source distribution
within 1 $\sigma$ total error (statistical one plus source selection).
We have checked that 27 source distributions out of 100 predict the
sufficient large-scale isotropy within 1 $\sigma$ statistical error.
In order to investigate the effects of the GMF on the large-scale
anisotropy, we also calculate the harmonic amplitude for the case of
$B_{\rm Solar}=0.0 \mu$ G.
For $B_{\rm Solar}=0.0 \mu$ G, the predicted arrival distribution is
relatively more isotropic than that for $B_{\rm Solar}=1.5 \mu$ G.
We also note that this tendency can be seen at lower energies ($\sim
10^{19}$ eV).
Because the deflection angle of cosmic rays with such energies by the
GMF is about $\sim {\rm a few} \times 10^{\circ}$, the harmonic
amplitude of arrival distribution of UHECRs can be affected by
anisotropy of the event distributions which is caused by the events
aligned according to the order of their energies, reflecting the
direction of the GMF.

\vspace{0.5cm}
%%%%%%%%%%%%%%%%%%%%%%%%%%%%%%%%%%%%%%%%%%%%%%%%%%%%%%%%%%%%%%%
\centerline{{\vbox{\epsfxsize=8.0cm\epsfbox{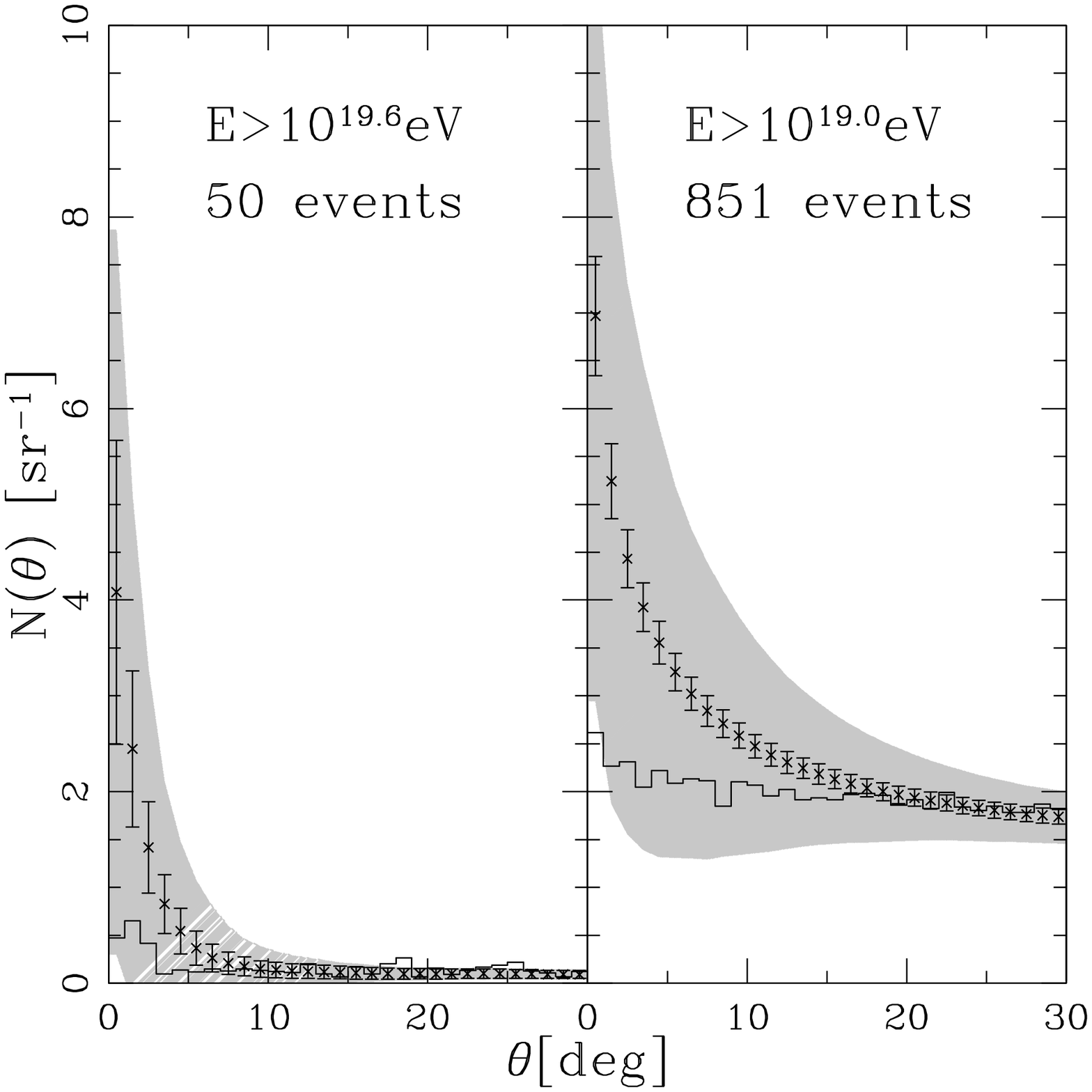}}}}
\figcaption{
Two point correlation function expected for our source model for $E>4
\times 10^{19}$ (left) and $E>10^{19}$ eV (right).
The errorbars represent the statistical fluctuations due to the finite
number of the simulated events, which is set to be equal to that
observed by the AGASA within $-10^{\circ} \le \delta \le 80^{\circ}$.
The shaded regions represent 1 $\sigma$ total error due to not only
the statistical error but also the source selections from our ORS
sample.
The histograms represent the AGASA data.
However, the AGASA data for $E>10^{19}$ eV are fitted to the result of
our calculation at larger angle ($30^{\circ}$), since we can not know
the normalization of the AGASA data with this energy.
\label{2P}}
\vspace{0.5cm}
%%%%%%%%%%%%%%%%%%%%%%%%%%%%%%%%%%%%%%%%%%%%%%%%%%%%%%%%%%%%%%%

In figure~\ref{2P}, we show two point correlation function predicted
by our source model for $E>4 \times 10^{19}$ (left) and $E>10^{19}$ eV
(right).
It is noted that we calculate two point correlation function for the
simulated events within only $-10^{\circ} \le \delta \le 80^{\circ}$
in order to compare with AGASA data.
The errorbars represent the statistical fluctuations due to the finite
number of the simulated events, which is set to be equal to that
observed by the AGASA (775 events for $E>10^{19}$ eV).
The shaded regions represent 1 $\sigma$ total error due to not only
the statistical error but also the source selections from our ORS
sample.
The event selections and the random source selections are performed
$20$ times and $100$ times, respectively.
The event numbers shown in this figures are averaged over all trials
of the event selections and the source selections.
The histograms represent the AGASA data.
However, the AGASA data for $E>10^{19}$ eV are fitted to the result of
our calculation at larger angle ($30^{\circ}$), since we can not know
the normalization of the AGASA data with this energy.

Clearly visible is that large peak at small angle scale is too strong
compared with the AGASA observation~\citep{takeda01}.
We have checked that when extremely nearby sources are selected by
accident, predicted small-scale anisotropy becomes to be very strong.
This is the reason for too large peak at small angle scale in
figure~\ref{2P}, where the averages and the variances are calculated
including such source distributions.

Provided that extremely nearby sources are selected by accident, not
only the small-scale anisotropy but also the large-scale isotropy is
inconsistent with the AGASA observation.
Accordingly, we calculate two point correlation function only for the
source distributions which predict the large-scale isotropy consistent
with uniform source distribution within 1 $\sigma$ statistical error.
The number of such source distributions is 27 out of all the 100
source selections.
The result is shown in figure~\ref{2P_1sig}.

Two point correlation function in figure~\ref{2P_1sig} exhibits a structure
that is similar to that seen in the AGASA data, that is, large peak at
small angle scale followed by a tail at large angles.
For $E>10^{19}$ eV, a peak at small angles is somewhat weaker than
that for $E>4 \times 10^{19}$ eV because of the large deflection by
the GMF.
This feature is also seen in AGASA data \citep*{takeda01}.
From this result, it can be understood that the source distributions
which predict sufficient large-scale isotropy also predict the
small-scale anisotropy that is similar to that seen in the AGASA data.

\vspace{0.5cm}
%%%%%%%%%%%%%%%%%%%%%%%%%%%%%%%%%%%%%%%%%%%%%%%%%%%%%%%%%%%%%%%
\centerline{{\vbox{\epsfxsize=8.0cm\epsfbox{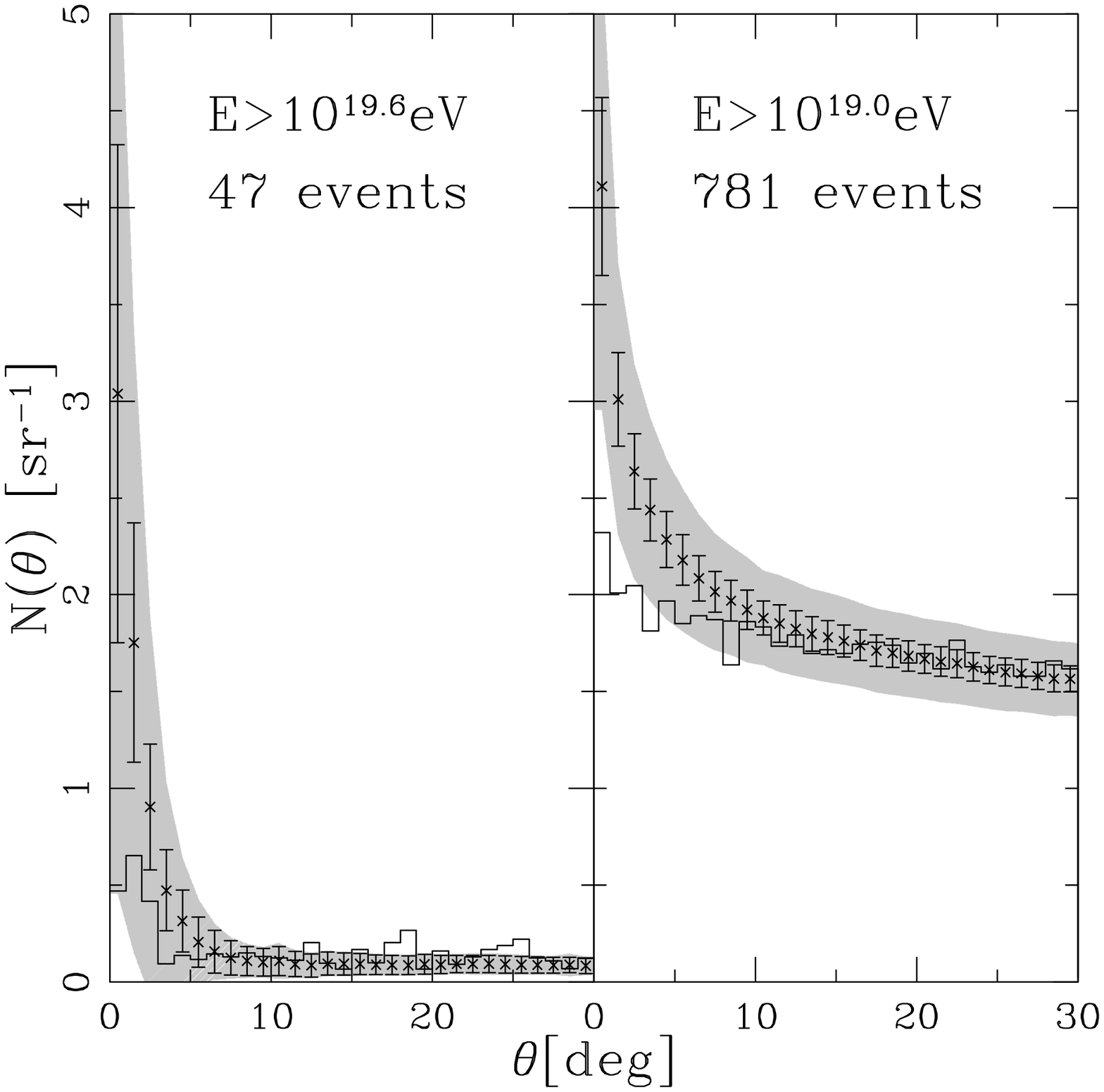}}}}
\figcaption{
Same as figure~\ref{2P}.
But, this is the result only for source distributions which predict
the large-scale anisotropy consistent with uniform source distribution
within 1 $\sigma$ statistical error.
\label{2P_1sig}}
\vspace{0.5cm}
%%%%%%%%%%%%%%%%%%%%%%%%%%%%%%%%%%%%%%%%%%%%%%%%%%%%%%%%%%%%%%%

However, we should note that the peak at small angle scale is still
relatively strong compared with the AGASA.
There may be two possible explanations for this fact.
First, we neglect the effects of the extragalactic magnetic field in
this study, in order to save the CPU time.
If we can include this effect by future studies, strong correlation at
small scale will be reduced because of the deflection of UHECRs in the
intergalactic space.
Second, we also neglect the random component of the GMF in order to
make easy to see the effect of the regular field.
This may also relax the large peak at small angle scale, provided that
there is the random component with same level of strength with the
regular component.
These issues are left for future investigations.

%-------------------------------------------------------------------------
\subsection{Future Prospects of UHECR arrival distribution}
\label{arrival_future}
%-------------------------------------------------------------------------

\begin{figure*}
\begin{center}
%\leavevmode
\epsscale{2.0} 
\plotone{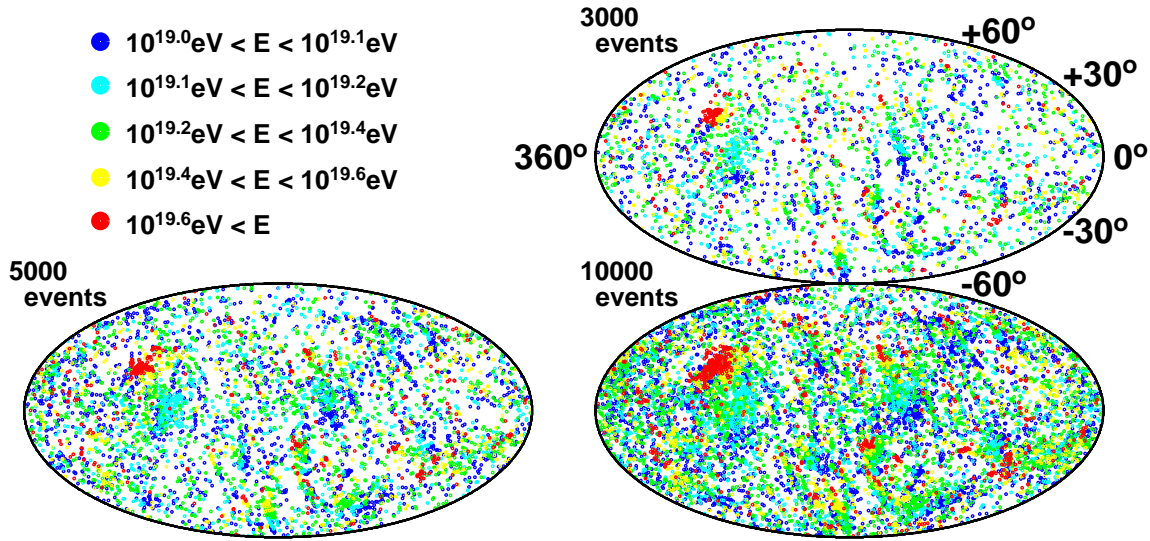} 
\caption{
Arrival directions of protons with $E>10^{19.0}$ eV at the
earth expected for the source model of
figure~\ref{source} in the galactic coordinate.
The events are shown by color according to their energies.
It is noted that the expected event rate by the Auger experiment is
$\sim 3000$ per year above $10^{19}$ eV.
\label{selected_many}}
\end{center}
\end{figure*}

In this subsection, we demonstrate the results of the UHECR arrival
distribution above $E>10^{19}$ eV with the event number expected by
future experiments, such as Auger, EUSO and OWL.
The results for the source model of figure~\ref{source} are shown in
figure~\ref{selected_many}.
The events are shown by color according to their energies.
It is noted that the expected event rate by the Auger experiment is
$\sim 3000$ per year above $10^{19}$ eV.
These results are extended versions of our previous study
\citep{yoshiguchi03c}, where we predict the UHECR arrival distribution
above $4 \times 10^{19}$ eV without modifications by the GMF.

Remarkable feature is the arrangement of clustered events at the
directions of nearby sources (see figure~\ref{source}).
The events are aligned according to the order of their energies
reflecting the direction of the GMF.
We will be able to obtain some kind of information about the GMF and
the chemical composition of UHECRs.
In forthcoming work, we plan studies about new statistical quantities
which allow us to obtain such invaluable information.

%%%%%%%%%%%%%%%%%%%%%%%%%%%%%%%%%%%%%%%%%%%%%%%%%
\section{SUMMARY AND DISCUSSION} \label{summary}
%%%%%%%%%%%%%%%%%%%%%%%%%%%%%%%%%%%%%%%%%%%%%%%%%

In this paper, we presented a new method for calculating the arrival
distribution of UHECRs, which can be applied to several source
location scenarios, including modifications by the GMF.
We performed numerical simulations of UHE anti-protons, which are
injected isotropically at the earth, in the Galaxy and recorded the
directions of velocities at the earth and outside the Galaxy for all
of the trajectories.
It is noted that we regard these anti-protons as PROTONs injected from
the outside of the Galaxy toward the earth.
We then selected some trajectories so that the resultant mapping of
the velocity directions outside the Galaxy of the selected
trajectories corresponds to our source location scenario, applying
Liouville's theorem.

There are two points of our improvement over the work of
\cite{stanev02}.
First, we calculated only the trajectories actually reaching the
detectors by propagating anti-protons backwards from Earth, instead of
propagating protons from the source and selecting those reaching the
Earth.
This helps us to save the CPU time efficiently and makes calculations
of propagation of cosmic rays even with lower energies ($\sim 10^{19}$
eV) possible within a reasonable time.
Second, we considered energy loss processes of UHE protons in the
intergalactic space, which is not taken into account in
\cite{stanev02}.
This enables us to include the effects of difference between resultant
energy spectra injected at different distances into numerical
calculations.
We can calculate the arrival distribution of UHECRs under more
realistic situation.

As an application of this method, we calculate the UHECR arrival
distribution above $10^{19}$ eV for the source model which is adopted
in our recent study (Paper I) and can explain the current AGASA
observation above $4 \times 10^{19}$ eV.
We found that the predicted large-scale anisotropy is fully consistent
with uniform source distribution, in the same manner as the current
AGASA data.
In order to investigate the effects of the GMF on the large-scale
anisotropy, we calculated the harmonic amplitude for the case of
$B_{\rm Solar}=0.0 \mu$ G.
For $B_{\rm Solar}=0.0 \mu$ G, the predicted arrival distribution is
relatively more isotropic than that for $B_{\rm Solar}=1.5 \mu$ G.
This would be due to anisotropy of the event distributions which is
caused by the events aligned according to the order of their energies,
reflecting the direction of the GMF.

It is also found that the calculated two point correlation function is
similar to that of AGASA data, when we restrict our attention to the
source distributions which predict sufficient isotropic arrival
distribution of UHECRs.
There may be effects of the extragalactic magnetic field and the random
component of the GMF on the large peak of two point correlation
function.
These issues are left for future studies.

Finally, we demonstrated the UHECR arrival distribution above
$10^{19}$ eV with the event number expected by future experiments in
the next few years.
The interesting feature of the resultant arrival distribution is the
events aligned according to the order of their energies,
reflecting the directions of the galactic magnetic field.
This is also pointed out by \cite{stanev02}.
This feature will become clear with increasing amount of data, and
allow us to obtain some kind of information about the composition of
UHECRs and the GMF.
In forthcoming work, we plan studies about new statistical quantities
which allow us to obtain such invaluable information.

In the present work, we calculate the arrival distribution of UHECRs
for our source location scenario which is adopted in our previous
study (Paper I).
However, it should be mentioned that the same results would be
obtained if the sources were truly drawn at random, as far as the
source number density is $\sim 10^{-6}$ Mpc$^{-3}$.
The results such as the events aligned according to the order of their
energies are independent on our assumption about the source
distribution.

In this study, we calculate the harmonic amplitude and two point
correlation function, which are only published quantities on UHECRs
observed by the AGASA including lower energy ones ($\sim 10^{19}$ eV).
In particular, the AGASA observation has published neither existence
nor non-existence of the events aligned according to the order of their
energies.
If more detailed event data with $E<4 \times 10^{19}$ eV are
published, we may obtain more strong constraint on our source model,
other than the source number density, using another statistical
quantities.
This is also one of future study plans.

%%%%%%%%%%%%%%%%%%%%%%%%%%%%%%%%%%%%%%%%%%%%%%%%%%%%%%%%%%%%%%%
%%%%%%%%%%%%%%%%%%%%%%%%%%%%%%%%%%%%%%%%%%%%%%%%%%%%%%%%%%%%%%%
\acknowledgments
%%%%%%%%%%%%%%%%%%%%%%%%%%%%%%%%%%%%%%%%%%%%%%%%%%%%%%%%%%%%%%%
%%%%%%%%%%%%%%%%%%%%%%%%%%%%%%%%%%%%%%%%%%%%%%%%%%%%%%%%%%%%%%%
The authors acknowledge fruitful discussions with S. Tsubaki.
This research was supported in part by Giants-in-Aid for Scientific
Research provided by the Ministry of Education, Science and Culture
of Japan through Research Grant No.S14102004 and No.14079202.

%%%%%%%%%%%%%%%%%%%%%%%%%%%%%%%%%%%%%%%%%%%%%%%%%%%%%%%%%%%%%%%%%%%%%%%
%\clearpage
%%%%%%%%%%%%%%%%%%%%%%%%%%%%%%%%%%%%%%%%%%%%%%%%%%%%%%%%%%%%%%%%%%%%%%%

%%%%%%%%%%%%%%%%%%%%%%%%%%%%%%%%%%%%%%%%%%%%%%%%%%%%%%%%%%%%%%%%%%%%%%%


\begin{thebibliography}{}
%\parskip=-1pt
%\baselineskip=14pt
%\bibitem[Bartelmann(1995)]{bartelmann95b}
%Bartelmann, M. 1995, \aap, 299, 11


\bibitem[Abu-Zayyad et al.(2002)]{abu02} Abu-Zayyad T. et al.
(The HiRes Collaboration) 2002, astro-ph/0208243

\bibitem[Alvarez-Muniz, Engel \& Stanev(2002)]{stanev02}
Alvarez-Muniz J., Engel R., \& Stanev T.
2002, \apj, 572, 185

\bibitem[Beck(2001)]{beck01}
Beck R. 2001, Space.\ Sci.\ Rev.\, 99, 243

\bibitem[Benson \& Linsley(1982)]{euso92} Benson R., \& Linsley J.
1982, \aap, 7, 161

\bibitem[Berezinsky \& Grigorieva(1988)]{berezinsky88}
Berezinsky V., \& Grigorieva S.I. 1988, \aap, 199, 1

\bibitem[Bird et al.(1994)]{bird94}
Bird D.J., et al. 1994, \apj, 424, 491

\bibitem[Capelle et al.(1998)]{capelle98} Capelle K.S., Cronin J.W.,
\& Parente G., Zas E. 1998, APh, 8, 321

\bibitem[Chodorowski, Zdziarske, \& Sikora(1992)]{chodorowski92}
Chodorowski M.J., Zdziarske A.A., \& Sikora M. 1992, \apj, 400, 181 

\bibitem[Cline \& Stecker(2000)]{owl00} Cline D.B., \& Stecker F.W.
OWL/AirWatch science white paper, astro-ph/0003459

%\bibitem[Fermi(1949)]{fermi49} Fermi E. 1949, Phys. Rev., 75, 1169

\bibitem[Greisen(1966)]{greisen66} Greisen K. 1966, Phys. Rev. Lett., 16, 748 

\bibitem[Han \& Qiao(1994)]{han94}
Han J.L., \& Qiao G.J. 1994, \aap, 288, 759

\bibitem[Han, Manchester, \& Qiao(1999)]{han99}
Han J.L., Manchester R.N., \& Qiao G.J. 1999,
MNRAS, 306, 371

\bibitem[Han(2001)]{han01}
Han J.L., 2001, Ap\&SS, 278, 181

\bibitem[Han(2002)]{han02}
Han J.L., 2002, astro-ph/0110319.

\bibitem[Hayashida et al.(1999)]{hayashida99} Hayashida N., et al. 1999
APh, 10, 303

\bibitem[Hayashida et al.(2000)]{hayashida00} Hayashida N., et al. 2000,
astro-ph/0008102

\bibitem[Ide et al.(2001)]{ide01} Ide Y., Nagataki S., Tsubaki S.,
Yoshiguchi H., \& Sato K. 2001, PASJ, 53, 1153

\bibitem[Isola \& Sigl(2002)]{isola02} Isola C., \& Sigl G. 2002,
astro-ph/0203273

\bibitem[Marco, Blasi, \& Olinto(2003)]{marco03} Marco D.D., Blasi P.,
\& Olinto A.V. 2003, astro-ph/0301497

\bibitem[Mucke et al.(2000)]{sophia00} Mucke A., Engel R., Rachen J.P.,
Protheroe R.J., \& Stanev T. 2000, Comput. Phys. Commun. 124, 290

\bibitem[Santiago et al.(1995)]{santiago95} Santiago B.X., Strauss M.A.,
Lahav O., Davis M., Dressler A., \& Huchra J.P. 1995, \apj, 446, 457

\bibitem[Sigl, Miniati, \& Ensslin(2003)]{sigl03} Sigl G., Miniati
F., \& Ensslin T.A. 2003, astro-ph/0302388

\bibitem[Stanev(1996)]{stanev96} Stanev T. 1996,
\apj, 479, 290

\bibitem[Takeda et al.(1998)]{takeda98} Takeda M., et al. 1998, Phys.
Rev. Lett., 81, 1163

\bibitem[Takeda et al.(1999)]{takeda99} Takeda M., et al. 1999, \apj, 522, 225

\bibitem[Takeda et al.(2001)]{takeda01} Takeda M., et al. 2001,
Proceeding of ICRC 2001, 341

\bibitem[Wilkinson et al.(1999)]{wilkinson99} Wilkinson C.R., et al. 1999,
APh,  12, 121

\bibitem[Yoshida \& Teshima(1993)]{yoshida93} Yoshida S., \& Teshima M.
1993, Prog. Theor. Phys. 89, 833

\bibitem[Yoshida et al.(1995)]{yoshida95}
Yoshida S., et al. 1995 
Astropart. Phys., 3, 105

\bibitem[Yoshiguchi et al.(2003a)]{yoshiguchi03a} Yoshiguchi H., Nagataki
S., Tsubaki S., \& Sato K. 2003a, \apj, 586, 1231 (Paper I,
astro-ph/0210132)

\bibitem[Yoshiguchi et al.(2003b)]{yoshiguchi03b} Yoshiguchi H., Nagataki
S., Sato K., Ohama N., \& Okamura S. 2003b, PASJ, 55, 121
(astro-ph/0212061)

\bibitem[Yoshiguchi, Nagataki, \& Sato(2003c)]{yoshiguchi03c}
Yoshiguchi H., Nagataki S., Sato K. 2003c, \apj, in press
(astro-ph/0302508)

%\bibitem[Zas(2001)]{zas01} Zas E. 2001, astro-ph/0103371

\bibitem[Zatsepin \& Kuz'min(1966)]{zatsepin66} Zatsepin G.T., \& Kuz'min V.A.
1966, JETP Lett., 4, 78
\end{thebibliography}
\end{document}